\documentclass[aps,prl,superscriptaddress,showpacs,showkeys,floatfix,twocolumn]{revtex4}

\usepackage{graphicx}	

\def\Journal#1#2#3#4{{#1}{\bf #2}, #3 (#4)}

\def\NIMA{{Nucl. Instrum. Methods}~{\bf A}}

\def\PLB{{Phys. Lett.}~{\bf B}}

\def\PRL{Phys. Rev. Lett.\ }

\def\PRC{{Phys. Rev.}~{\bf C}}
\def\ZPC{{Z. Phys.}~{\bf C}}

\begin{document}

\title{Deuteron and antideuteron production 
in Au+Au collisions at $\sqrt{s_{_{NN}}}~=~200$~GeV}

\newcommand{\abilene}{Abilene Christian University, Abilene, TX 79699, USA}
\newcommand{\acadsin}{Institute of Physics, Academia Sinica, Taipei 11529, Taiwan}
\newcommand{\banaras}{Department of Physics, Banaras Hindu University, Varanasi 221005, India}
\newcommand{\barc}{Bhabha Atomic Research Centre, Bombay 400 085, India}
\newcommand{\bnl}{Brookhaven National Laboratory, Upton, NY 11973-5000, USA}
\newcommand{\caucr}{University of California - Riverside, Riverside, CA 92521, USA}
\newcommand{\ciae}{China Institute of Atomic Energy (CIAE), Beijing, People's Republic of China}
\newcommand{\cns}{Center for Nuclear Study, Graduate School of Science, University of Tokyo, 7-3-1 Hongo, Bunkyo, Tokyo 113-0033, Japan}
\newcommand{\columbia}{Columbia University, New York, NY 10027 and Nevis Laboratories, Irvington, NY 10533, USA}
\newcommand{\dapnia}{Dapnia, CEA Saclay, F-91191, Gif-sur-Yvette, France}
\newcommand{\debrecen}{Debrecen University, H-4010 Debrecen, Egyetem t{\'e}r 1, Hungary}
\newcommand{\fsu}{Florida State University, Tallahassee, FL 32306, USA}
\newcommand{\gsu}{Georgia State University, Atlanta, GA 30303, USA}
\newcommand{\hiroshima}{Hiroshima University, Kagamiyama, Higashi-Hiroshima 739-8526, Japan}
\newcommand{\ihepprot}{Institute for High Energy Physics (IHEP), Protvino, Russia}
\newcommand{\isu}{Iowa State University, Ames, IA 50011, USA}
\newcommand{\jinrdubna}{Joint Institute for Nuclear Research, 141980 Dubna, Moscow Region, Russia}
\newcommand{\kaeri}{KAERI, Cyclotron Application Laboratory, Seoul, South Korea}
\newcommand{\kangnung}{Kangnung National University, Kangnung 210-702, South Korea}
\newcommand{\kek}{KEK, High Energy Accelerator Research Organization, Tsukuba-shi, Ibaraki-ken 305-0801, Japan}
\newcommand{\kfki}{KFKI Research Institute for Particle and Nuclear Physics (RMKI), H-1525 Budapest 114, POBox 49, Hungary}
\newcommand{\korea}{Korea University, Seoul, 136-701, Korea}
\newcommand{\kurchatov}{Russian Research Center ``Kurchatov Institute", Moscow, Russia}
\newcommand{\kyoto}{Kyoto University, Kyoto 606, Japan}
\newcommand{\labllr}{Laboratoire Leprince-Ringuet, Ecole Polytechnique, CNRS-IN2P3, Route de Saclay, F-91128, Palaiseau, France}
\newcommand{\lawllnl}{Lawrence Livermore National Laboratory, Livermore, CA 94550, USA}
\newcommand{\losalamos}{Los Alamos National Laboratory, Los Alamos, NM 87545, USA}
\newcommand{\lpc}{LPC, Universit{\'e} Blaise Pascal, CNRS-IN2P3, Clermont-Fd, 63177 Aubiere Cedex, France}
\newcommand{\lund}{Department of Physics, Lund University, Box 118, SE-221 00 Lund, Sweden}
\newcommand{\muenster}{Institut f\"ur Kernphysik, University of Muenster, D-48149 Muenster, Germany}
\newcommand{\myongji}{Myongji University, Yongin, Kyonggido 449-728, Korea}
\newcommand{\nagasaki}{Nagasaki Institute of Applied Science, Nagasaki-shi, Nagasaki 851-0193, Japan}
\newcommand{\newmex}{University of New Mexico, Albuquerque, NM, USA}
\newcommand{\nmsu}{New Mexico State University, Las Cruces, NM 88003, USA}
\newcommand{\ornl}{Oak Ridge National Laboratory, Oak Ridge, TN 37831, USA}
\newcommand{\orsay}{IPN-Orsay, Universite Paris Sud, CNRS-IN2P3, BP1, F-91406, Orsay, France}
\newcommand{\pnpi}{PNPI, Petersburg Nuclear Physics Institute, Gatchina, Russia}
\newcommand{\riken}{RIKEN (The Institute of Physical and Chemical Research), Wako, Saitama 351-0198, JAPAN}
\newcommand{\rkrbrc}{RIKEN BNL Research Center, Brookhaven National Laboratory, Upton, NY 11973-5000, USA}
\newcommand{\saispbstu}{St. Petersburg State Technical University, St. Petersburg, Russia}
\newcommand{\saopaulo}{Universidade de S{\~a}o Paulo, Instituto de F\'{\i}sica, Caixa Postal 66318, S{\~a}o Paulo CEP05315-970, Brazil}
\newcommand{\seoulnat}{System Electronics Laboratory, Seoul National University, Seoul, South Korea}
\newcommand{\stonybrkc}{Chemistry Department, Stony Brook University, SUNY, Stony Brook, NY 11794-3400, USA}
\newcommand{\stonycrkp}{Department of Physics and Astronomy, Stony Brook University, SUNY, Stony Brook, NY 11794, USA}
\newcommand{\subatech}{SUBATECH (Ecole des Mines de Nantes, CNRS-IN2P3, Universit{\'e} de Nantes) BP 20722 - 44307, Nantes, France}
\newcommand{\tenn}{University of Tennessee, Knoxville, TN 37996, USA}
\newcommand{\titech}{Department of Physics, Tokyo Institute of Technology, Tokyo, 152-8551, Japan}
\newcommand{\tsukuba}{Institute of Physics, University of Tsukuba, Tsukuba, Ibaraki 305, Japan}
\newcommand{\vandy}{Vanderbilt University, Nashville, TN 37235, USA}
\newcommand{\waseda}{Waseda University, Advanced Research Institute for Science and Engineering, 17 Kikui-cho, Shinjuku-ku, Tokyo 162-0044, Japan}
\newcommand{\weizmann}{Weizmann Institute, Rehovot 76100, Israel}
\newcommand{\yonsei}{Yonsei University, IPAP, Seoul 120-749, Korea}
\affiliation{\abilene}
\affiliation{\acadsin}
\affiliation{\banaras}
\affiliation{\barc}
\affiliation{\bnl}
\affiliation{\caucr}
\affiliation{\ciae}
\affiliation{\cns}
\affiliation{\columbia}
\affiliation{\dapnia}
\affiliation{\debrecen}
\affiliation{\fsu}
\affiliation{\gsu}
\affiliation{\hiroshima}
\affiliation{\ihepprot}
\affiliation{\isu}
\affiliation{\jinrdubna}
\affiliation{\kaeri}
\affiliation{\kangnung}
\affiliation{\kek}
\affiliation{\kfki}
\affiliation{\korea}
\affiliation{\kurchatov}
\affiliation{\kyoto}
\affiliation{\labllr}
\affiliation{\lawllnl}
\affiliation{\losalamos}
\affiliation{\lpc}
\affiliation{\lund}
\affiliation{\muenster}
\affiliation{\myongji}
\affiliation{\nagasaki}
\affiliation{\newmex}
\affiliation{\nmsu}
\affiliation{\ornl}
\affiliation{\orsay}
\affiliation{\pnpi}
\affiliation{\riken}
\affiliation{\rkrbrc}
\affiliation{\saispbstu}
\affiliation{\saopaulo}
\affiliation{\seoulnat}
\affiliation{\stonybrkc}
\affiliation{\stonycrkp}
\affiliation{\subatech}
\affiliation{\tenn}
\affiliation{\titech}
\affiliation{\tsukuba}
\affiliation{\vandy}
\affiliation{\waseda}
\affiliation{\weizmann}
\affiliation{\yonsei}
\author{S.S.~Adler}	\affiliation{\bnl}
\author{S.~Afanasiev}	\affiliation{\jinrdubna}
\author{C.~Aidala}	\affiliation{\bnl}
\author{N.N.~Ajitanand}	\affiliation{\stonybrkc}
\author{Y.~Akiba}	\affiliation{\kek} \affiliation{\riken}
\author{J.~Alexander}	\affiliation{\stonybrkc}
\author{R.~Amirikas}	\affiliation{\fsu}
\author{L.~Aphecetche}	\affiliation{\subatech}
\author{S.H.~Aronson}	\affiliation{\bnl}
\author{R.~Averbeck}	\affiliation{\stonycrkp}
% ...
\collaboration{PHENIX Collaboration} \noaffiliation

\date{\today}

\begin{abstract}
The production of deuterons and antideuterons in the transverse 
momentum range \mbox{1.1$\ <\ p_T\ <\ $4.3}~GeV/$c$ at mid-rapidity
in Au + Au collisions at $\sqrt{s_{NN}} = 200$ GeV has been studied by the
PHENIX experiment at RHIC.  A coalescence analysis comparing the deuteron
and antideuteron spectra with those of protons and antiprotons, has been
performed. The coalescence probability is equal for both deuterons and
antideuterons and increases as a function of $p_T$, which is consistent
with an expanding collision zone. Comparing (anti)proton yields :
$\overline{p}/p~=~0.73\pm0.01$, with (anti)deuteron yields:  
$\overline{d}/d~=~0.47\pm0.03$, we estimate that
$\bar{n}/n~=~0.64\pm0.04$.

\end{abstract}



\maketitle

Ultrarelativistic heavy ion collisions are used to study the behavior of
nuclear matter at extreme conditions of temperature and density, similar
to those that existed in the universe a few microseconds after the Big
Bang.  Previous measurements indicate that high particle
multiplicities~\cite{mult1} and large $\bar{p}/p$ ratios prevail at the
Relativistic Heavy-Ion Collider (RHIC), which is expected for a nearly net
baryon free region~\cite{ppg006}. As the hot, dense system of particles
cools, it expands and the mean free path increases until the particles
cease interacting (``freezeout''). At this point, light nuclei like
deuterons and antideuterons ($d$ and $\bar{d}$) can be formed, with a
probability proportional to the product of the phase space densities of
their constituent nucleons~\cite{csernai,mekjian}.  Thus, invariant yield of
deuterons, compared to the protons~\cite{ppg009,ppg026} from which they
coalesce, provides information about the size of the emitting system and
its space-time evolution.

PHENIX~\cite{phenixnim} at RHIC, is a versatile detector designed to study
the production of leptons, photons, and hadrons over a wide momentum
range. In this Letter, results on $d$ and $\bar{d}$ production in Au+Au
interactions at $\sqrt{s_{NN}} = 200$ GeV are presented. For the sake of
brevity, in the rest of this Letter, our statements will generally apply
to both particles and antiparticles.

The East central tracking spectrometer in the PHENIX
detector~\cite{ppg009,phenixnim,centnim} is used in this analysis. The
information from the PHENIX Beam-Beam Counters (BBC) and Zero-Degree
Calorimeters (ZDC) is used for triggering and event selection. The BBCs
are {\v C}erenkov-counters surrounding the beam pipe in the pseudorapidity
interval $3.0<|\eta|<3.9$, and provide the start timing signal. The ZDCs
are hadronic calorimeters 18~m downstream of the interaction region and
detect spectator neutrons in a narrow forward cone. Particle
identification in the central rapidity region is achieved by measuring
momentum (by drift chamber) and time of flight (by time-of-flight
detector). The drift chamber (DC) and two layers of pad chambers (PC) are
used for tracking and momentum reconstruction~\cite{centnim}. The
time-of-flight detector (TOF) spans the pseudorapidity range $|\eta|<0.35$
and $\Delta \phi = \pi/4$ azimuthally.  The TOF consists of plastic
scintillators, with a combined time resolution of $\approx$ 115 ps. The
TOF thus provides identification of $d$ and $\bar{d}$ in the transverse
momentum ($p_T$) range $1.1<p_T<4.3$~GeV/$c$. For $p_T<1.1$~GeV/$c$, the
signal to background ratio suffers due to multiple scattering and energy
loss effects.

\begin{figure}[thb]
\includegraphics[width=1.0\linewidth]{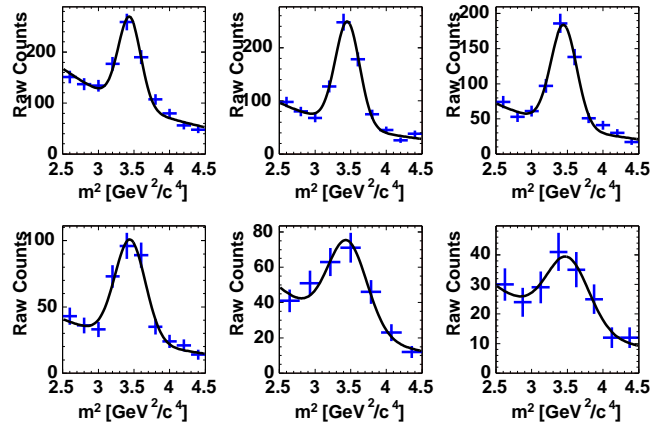}  
\caption{\label{fig:ppg020_fig1} (color online)
Histograms of the mass squared for identified 
antideuterons in the transverse momentum range 1.1 $<p_T<$ 3.5 
GeV/$c$ (in 400 MeV/$c$ increments), with Gaussian fits including 
an exponential background.}
\end{figure}

The dataset for this analysis includes 21.6 million minimum bias events.
The minimum bias cross section corresponds to $92.2^{+2.5}_{-3}$\% of the
total inelastic Au+Au cross section (6.9 b)~\cite{ppgmb}. Using the
momentum determined by the DC, which has a resolution of 
$\delta p/p \approx 0.7\% \oplus 1\% p$ GeV/$c$, and the time of flight
from the event vertex provided by the TOF, the mass of the particle is
determined. The $d$ and $\bar{d}$ yields are obtained by fitting the mass
squared distributions to the sum of a Gaussian signal and an exponential
background. Examples of mass squared distributions with fits for
antideuterons in minimum bias collisions are shown in
Fig.~\ref{fig:ppg020_fig1}.

The raw yields are corrected for effects of detector acceptance,
reconstruction efficiency and detector occupancy.  Corrections are
determined by reconstructing single deuterons simulated using
GEANT~\cite{geant} and a detector response model of PHENIX, using the
method described in~\cite{ppg026}. The track reconstruction efficiency
decreases in high multiplicity events because of high detector occupancy.
This effect can be slightly larger for slower, heavier particles, due to
detector dead times between successive hits. Occupancy effects on
reconstruction efficiency ($\approx$ 83.5\% for 0-20\% most central
events) are evaluated by embedding simulated single particle Monte Carlo
events in real events. Since the hadronic interactions of nuclei are not
treated by GEANT, a correction needs to be applied to account for the
hadronic absorption of $d$ and $\bar{d}$ (including annihilation). The
$d$- and $\bar{d}$-nucleus cross sections are calculated from
parameterizations of the nucleon and anti-nucleon cross
sections:
\begin{equation}
\sigma_{d/\bar{d},A} =  [\sqrt{\sigma_{N/\bar{N},A}} +  \Delta_d ]^{2}
\label{eq:dbarcs}
\end{equation}

The limited data available on deuteron induced interactions~\cite{jaros}
indicate that the term $\Delta_d$ is independent of the nuclear mass
number $A$ and that $\Delta_d = 3.51\pm 0.25$~mb$^{1/2}$. The hadronic
absorption varies only slightly over the applicable $p_T$ range and is
$\approx$ 10\% for $d$ and $\approx$ 15\% for $\bar{d}$. 
The background contribution from deuterons knocked out due to the 
interaction of the produced particles with the beam pipe is estimated 
using simulations and found to be negligible in the momentum range of our 
measurement.

\begin{figure}[thb]
\includegraphics[width=1.0\linewidth]{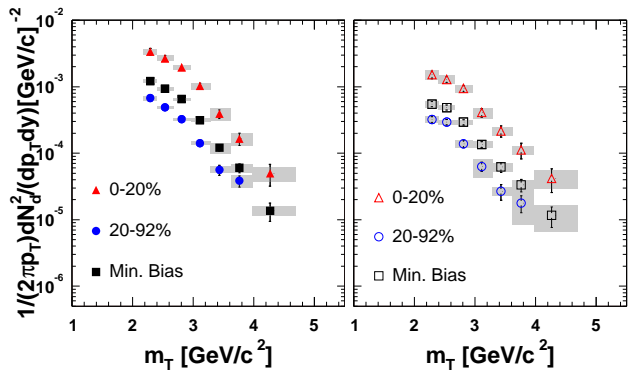}  
\caption{\label{fig:ppg020_fig2} (color online)
Corrected spectra for deuterons (left panel) and anti-deuterons 
(right panel) for different centralities are plotted vs $m_T$. Error bars
indicate statistical errors and grey bands the systematic errors. Values are 
plotted at the ``true'' mean value of $m_T$ of each 
bin, the extent of which is indicated by the width of the grey bars along 
x-axis.}
\end{figure}

Figure~\ref{fig:ppg020_fig2} shows the corrected $d$ and $\bar{d}$ invariant
yields as a function of transverse mass ($m_T$ in the range $1.1<p_T<4.3$
GeV/$c$, for minimum bias events, and two centrality bins: 0-20\% (most
central), 20-92\% (non-central). The 20-92\% centrality bin is dominated
by mid-central events, due to larger track multiplicities relative to
peripheral events.

Systematic uncertainties have several sources: errors in particle
identification, DC-TOF hit match efficiency, the uncertainty in momentum
scale, $d$ and $\bar{d}$ hadronic interaction correction, and uncertainty
in occupancy corrections. All the systematic uncertainties are added in
quadrature, depicted by grey bars in Fig.~\ref{fig:ppg020_fig2}.

\begin{table}
\caption{\label{tab:teff} 
The inverse slope parameter $T_{eff}$ obtained from a $m_T$
exponential fit to the spectra along with multiplicity $dN/dy$ and mean 
transverse momentum $\langle p_T \rangle$ obtained from a Boltzman 
distribution for different centralities:}
\begin{ruledtabular}\begin{tabular}{lll}
\textbf{$T_{eff}$ [MeV]}  & \textbf{Deuterons} & \textbf{Anti-deuterons} \\
\hline
Minimum Bias&519 $\pm$ 27	&512 $\pm$ 32\\
0-20\%	  &536 $\pm$ 32	&562 $\pm$ 51\\
20-92\%  &475 $\pm$ 29	&456 $\pm$ 35\\
\hline
\textbf{$dN/dy$}  &	  &	    \\
\hline
Minimum Bias&0.0250 $\pm^{0.0006(stat.)}_{0.005(sys.)}$	
&0.0117 $\pm^{0.0003(stat.)}_{0.002(sys.)}$\\
0-20\% &0.0727 $\pm^{0.0022(stat.)}_{0.0141(sys.)}$
&0.0336 $\pm^{0.0013(stat.)}_{0.0057(sys.)}$\\
20-92\% &0.0133 $\pm^{0.0004(stat.)}_{0.0029(sys.)}$	
&0.0066 $\pm^{0.0002(stat.)}_{0.0015(sys.)}$\\
\hline
\textbf{$\langle p_T \rangle$ [GeV/$c$]}&	  &	    \\
\hline
Minimum Bias&1.54 $\pm^{0.04(stat.)}_{0.13(sys.)}$
&1.52 $\pm^{0.05(stat.)}_{0.12(sys.)}$\\
0-20\%&1.58 $\pm^{0.05(stat.)}_{0.13(sys.)}$	
&1.62 $\pm^{0.07(stat.)}_{0.1(sys.)}$\\
20-92\%&1.45 $\pm^{0.05(stat.)}_{0.15(sys.)}$	
&1.41 $\pm^{0.06(stat.)}_{0.15(sys.)}$
\end{tabular} \end{ruledtabular}
\end{table}

The $p_T$ spectra $Ed^3N/d^3p$ are fitted in the range
$1.1<p_T<3.5$~GeV/$c$ to an exponential distribution in
$m_T=\sqrt{p_T^2+m^2}$. The inverse slopes ($T_{eff}$) of the spectra are
tabulated in Table~\ref{tab:teff}. The deuteron inverse slopes of
$T_{eff}$ = 500--520 MeV are considerably higher than the $T_{eff}$ =
300--350 MeV observed for protons~\cite{ppg009,ppg026}.  The invariant
yields and the average transverse momenta ($\langle p_T \rangle$) are
obtained by summing the data over $p_T$ and using a Boltzmann
distribution: $\frac{d^2N}{2\pi m_Tdm_Tdy} \propto m_Te^{-m_T/T_{eff}}$,
(which gives a slightly better $\chi^2/n.d.f.=4.8/3$ vs. $\chi^2/n.d.f.=5.6/3$
for exponential fit) to extrapolate to low $m_T$ regions where we have no
data. The extrapolated yields constitute $\approx$ 42\% of our total
yields.  The rapidity distributions, $dN/dy$, and the mean transverse
momenta, $\langle p_T \rangle$, are compiled in Table~\ref{tab:teff} for
three different centrality bins. Systematic uncertainties on $dN/dy$ and
$\langle p_T \rangle$ are estimated by using an exponential in $p_T$ and a
``truncated'' Boltzman distribution (assumed flat for $p_T < 1.1$ GeV/$c$)
for alternative extrapolations.

With a binding energy of 2.24 MeV, the deuteron is a very loosely bound
state. Thus, it is formed only at a later stage in the collision, most
likely after elastic hadronic interactions have ceased; the proton and
neutron must be close in space and tightly correlated in velocity to
coalesce. As a result, $d$ and $\bar{d}$ yields are a sensitive measure of
correlations in phase space and can provide information about the
space-time evolution of the system. If deuterons are formed by coalescence
of protons and neutrons, the invariant deuteron yield can be
related~\cite{butler} to the primordial nucleon yields by:
\begin{equation}
E_d\frac{d^3N_d}{d^3p_d}\biggr|_{p_d=2p_p} = B_2\left(E_p\frac{d^3N_p}{d^3p_p}\right)^2
\label{eq:coal}
\end{equation}
where $B_2$ is the coalescence parameter, with the subscript implying that
two nucleons are involved in the coalescence. The above equation includes
an implicit assumption that the ratio of neutrons to protons is unity. The
proton and antiproton spectra~\cite{ppg026} are corrected for feed-down
from $\Lambda$ and $\bar{\Lambda}$ decays by using a MC simulation tuned
to reproduce the particle ratios: ($\Lambda/p$ and
$\bar{\Lambda}/\bar{p}$) measured by PHENIX at 130 GeV~\cite{lambda130}.

\begin{figure}[thb]
\includegraphics[width=1.0\linewidth]{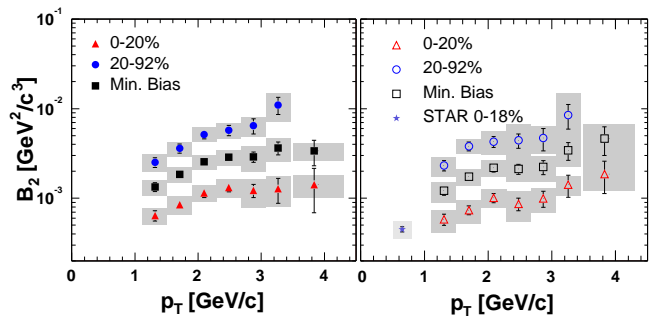}  
\caption{\label{fig:ppg020_fig3} (color online)
Coalescence parameter $B_2$ vs $p_T$ for deuterons (left panel) and anti-deuterons (right panel).
Grey bands indicate the 
systematic errors. Values are plotted at the ``true'' mean value of $p_T$ of 
each bin, the extent of which is indicated by the width of the grey bars along
x-axis.}
\end{figure}

Figure~\ref{fig:ppg020_fig3} displays the coalescence parameter $B_2$ as a
function of $p_T$ for different centralities.  The decreased $B_2$ in more
central collisions implies that in larger sources, the average relative
separation between nucleons increases, thus decreasing the probability of
formation of deuterons.  We also observe that $B_2$ increases with $p_T$. This
is consistent with an expanding source because position-momentum correlations
lead to a higher coalescence probability at larger $p_T$.  The $p_T$-dependence
of $B_2$ can also provide information about the density profile of the source as
well as the expansion velocity distribution. It has been
shown~\cite{heinz_prc99} that generally a Gaussian source density profile leads
to a constant $B_2$ with $p_T$ as it gives greater weight to the center of the
system, where radial expansion is weakest.  This is not supported by our data,
which shows a rise in $B_2$ with $p_T$.

Thermodynamic models~\cite{mekjian} predict that $B_2$ scales with the
inverse of the effective volume $V_{eff}$ ($B_2 \propto 1/V_{eff}$).  The
$d$ and $\bar{d}$ spectra are affected by radial flow, which concentrates
the coalescing protons and neutrons, affecting phase space correlations,
thereby limiting the applicability of a simple thermodynamical model to
determine an effective source size.

\begin{figure}[thb]
\includegraphics[width=1.0\linewidth]{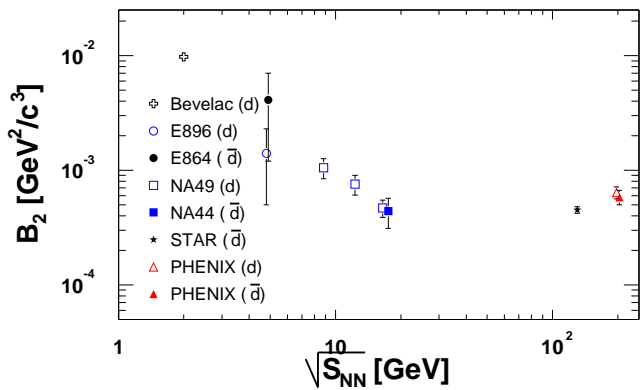}  
\caption{\label{fig:ppg020_sqsplot} (color online)
Comparison of the coalescence parameter for deuterons
and anti-deuterons ($p_T$ = 1.3 GeV/$c$) with other experiments at different 
values of $\sqrt{s}$.}
\end{figure}

Figure~\ref{fig:ppg020_sqsplot} compares $B_2$ for most central 
collisions to results at lower 
$\sqrt{s}$~\cite{eos,e896,e864,na49prc,na44prl,star}. 
Note that $B_2$ is nearly independent of $\sqrt{s}$, 
indicating that the source volume does not change appreciably with 
center-of-mass energy (with the caveat that $B_2$ 
varies as a function of $p_T$, centrality and rapidity). Similar behavior is
seen for $B_2$ for deuterons~\cite{na49prc} as a function of $\sqrt{s}$. 
This observation is consistent with what has been observed in Bose-Einstein 
correlation Hanbury-Brown Twiss (HBT) analysis at RHIC~\cite{hbt130}
for identified particles. The coalescence parameter $B_2$ for $d$ and 
$\bar{d}$, is equal within errors, indicating that nucleons and 
antinucleons have the same temperature, flow and freeze-out 
density distributions. 

The ratio $\bar{n}/n$ can be estimated from the data based on the 
thermal chemical model. Assuming thermal and chemical equilibrium, 
the chemical fugacities are determined from the particle/anti-particle 
ratios~\cite{heinz_prc99}:
\begin{equation}
\frac{E_A(d^3N_A/d^3p_A)}{E_{\bar{A}}(d^3N_{\bar{A}}/d^3p_{\bar{A}})} = 
\exp\left(\frac{2\mu_A}{T}\right) = \lambda_A^2
\label{eq:fugacity}
\end{equation}

\begin{figure}[thb]
\includegraphics[width=1.0\linewidth]{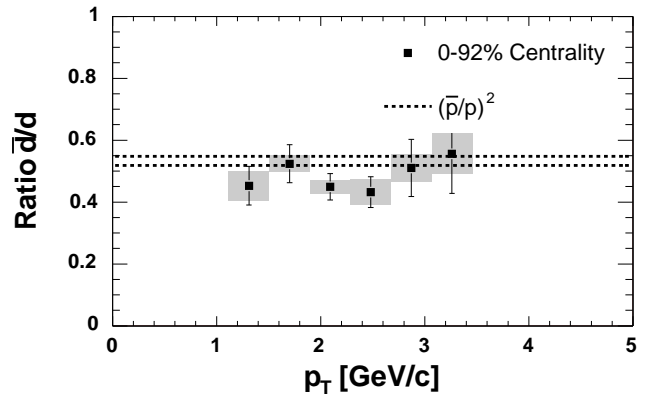}  
\caption{\label{fig:ppg020_dbard}
$\bar{d}/d$ ratio vs. $p_T$ for minimum bias data. The dashed lines 
represent the square of the measured $\bar{p}/p$ ratio as a function of $p_T$ 
within uncertainties.}
\end{figure}

Figure~\ref{fig:ppg020_dbard} shows that the $\bar{d}/d$ ratio 
is independent of centrality, 
and $p_T$ within errors. The average value of $\bar{d}/d$ is 0.47 
$\pm$ 0.03, consistent with the square of the ratio
$\bar{p}/p = 0.73 \pm 0.01$~\cite{ppg026} within statistical and systematic 
uncertainties. This is expected if deuterons are formed by 
coalescence of comoving nucleons and $\bar{p}/p = 
\bar{n}/n$. Using the ratio $p/\bar{p}$, the extracted proton fugacity 
is $\lambda_p = \exp(\mu_p/T)$ = 1.17 $\pm$ 0.01.
Similarly, using the $d/\bar{d}$ ratio, the extracted deuteron fugacity is 
$\lambda_d = \exp[(\mu_p+\mu_n)/T]$ = 1.46 $\pm$ 0.05.
From this, the neutron fugacity can be estimated to be
$\lambda_n = \exp(\mu_n/T)$ = 1.25 $\pm$ 0.04, which results in 
$\bar{n}/n$ = 0.64 $\pm$ 0.04.
These estimates, along with equality of $B_2$ for $d$ and $\bar{d}$ 
indicate that, within errors, $\mu_n \geq \mu_p$. This is 
expected since the entrance Au+Au channel has larger 
net neutron density than net proton density. 

To summarize, the transverse momentum spectra of $d$ and $\bar{d}$ in the
range $1.1<p_T<4.3$~GeV/$c$, have been measured at mid-rapidty in Au+Au
collisions at $\sqrt{s_{NN}}$ = 200 GeV, and are found to be less steeply
falling than
proton (and antiproton) spectra. This behavior is consistent with a constant
(flat) source density profile. The extracted coalescence parameter $B_2$
increases with $p_T$, which is indicative of an expanding source. $B_2$
decreases for more central collisions, consistent with an increasing
source size with centrality.  The $B_2$ measured in nucleus-nucleus
collisions is independent of $\sqrt{s_{NN}}$ above 12 GeV, 
consistent with Bose-Einstein correlation measurements of the radii of the
source.  $B_2$ is equal within errors for both deuterons and
anti-deuterons.  From the measurements, it is estimated that 
$\bar{n}/n$ = 0.64 $\pm$ 0.04.



We thank the staff of the Collider-Accelerator and Physics
Departments at BNL for their vital contributions.  We acknowledge
support from the Department of Energy and NSF (U.S.A.), MEXT and
JSPS (Japan), CNPq and FAPESP (Brazil), NSFC (China), CNRS-IN2P3
and CEA (France), BMBF, DAAD, and AvH (Germany), OTKA (Hungary), 
DAE and DST (India), ISF (Israel), KRF and CHEP (Korea),
RMIST, RAS, and RMAE, (Russia), VR and KAW (Sweden), U.S. CRDF 
for the FSU, US-Hungarian NSF-OTKA-MTA, and US-Israel BSF.


\clearpage

\end{document}